\documentstyle[aps,pre]{revtex}
\newcommand{\be}{\begin{equation}}
\newcommand{\ee}{\end{equation}}
\newcommand{\clH}{{\cal H}}
\newcommand{\clF}{{\cal F}}

\newcommand{\clM}{{\cal M}}

\newcommand{\bea}{\begin{eqnarray}}
\newcommand{\eea}{\end{eqnarray}}

\newcommand{\prt}{\partial}
\newcommand{\thh}{\tau_{\hbar}}
\newcommand{\rgl}{\rangle}
\newcommand{\lgl}{\langle}
\newcommand{\ep}{\epsilon}

\begin{document}

\title{Loschmidt echo for a chaotic oscillator}
\author{A. Iomin \\
Department of Physics, Technion, Haifa, 32000, Israel.  }

\maketitle

\begin{abstract}

Chaotic dynamics of a nonlinear oscillator is 
considered in the semiclassical approximation. 
The Loschmidt echo is calculated for a time scale which is of the power 
law in semiclassical parameter. It is shown that an exponential decay of the 
Loschmidt echo is due to a Lyapunov exponent and it has a pure classical 
nature.


\end{abstract}

Classical chaotic dynamics can be characterized by a Lyapunov
exponent $\Lambda$. The quantized procedure stops the classical spread
of stretching and folding due to the uncertainty principle and, as a
result, breaks the applicability of semiclassical approximation. The
corresponding breaking time was found in \cite{bz} to be:
\be\label{eqs1}
\tau_{\hbar} = (1/\Lambda ) \ln (I_0 /\hbar ),
\ee
where $I_0$ is a characteristic action. It indicates a fast (exponential)
growth of quantum corrections to the classical dynamics due to chaos.
Recently this result gained a renewed interest also
related  to the fidelity of wave functions or the Loschmidt echo. 
In the field of quantum chaos a question on stability of trajectories can 
be re-addressed to stability of wave functions respect to a small 
variation of a control parameter
\cite{perez} called the fidelity of the wave functions, which is a 
measure of a quantum reversibility. It is also named
``Loschmidt echo'' \cite{jpprl01}, and it is known that on the time scale
of the order of $\thh$ it decays exponentially $\sim e^{-\Lambda t}$ 
\cite{jpprl01}.
Different regimes of decay of the Loschmidt echo have been observed for 
{\em i.g.}
integrable \cite{eckhardt,beneti}, chaotic \cite{eckhardt,stb03,jiri} 
and pseudo-integrable \cite{jacq} systems (see also references therein).

In the present paper, we show that the Loschmidt echo in the nonlinear
kicked oscillator decays exponentially due to the Lyapunov exponent in a, 
so--called, Lyapunov regime \cite{jpprl01,jiri}. 
An analytical expression for this behavior in the framework of 
the semiclassical expansion is obtained. 

Dynamics of a nonlinear oscillator with the Hamiltonian 
$\clH_0=\hbar\omega a^{\dag}a+\hbar^2\mu(a^{\dag}a)^2 $ is integrable 
\cite{biz81}. Here $\omega$ and $\mu$ are linear frequency and nonlinearity 
parameters correspondingly, while, for the chosen notation, annihilation and 
creation operators have the commutation rule $[a,a^{\dag}]=1$.
There is nontrivial semiclassical expansion that leads to an appearance of 
so--called $D$--forms.
These forms are determined as derivatives over the initial conditions,
say, $(\alpha,\alpha^*)$ as
\be\label{eqs2}
D\equiv D(A,B)=\left(\frac{\prt A}{\prt\alpha}\right)
\cdot\left(\frac{\prt B}{\prt\alpha^*}\right) \, ,
\ee
where $A(B)=\lgl\psi(\alpha)|\hat{A}(\hat{B})|\psi(\alpha)\rgl$ is the
average value of the operator. 
Therefore, the $D$-forms determine the local instability of a dynamical
system. In the presence of a perturbation dynamics becomes chaotic
and the exponential growth of the $D$--forms due to the Lyapunov 
exponents leads to the logarithmic breaking time of Eq. 
(\ref{eqs1}), known also as the Ehrenfest time. That result could be 
transparently seen in the 
coherent state basis \cite{bz,zasl,biz81}, and it is independent of the 
choice of the initial basis of wave functions \cite{bi}.
Role of the $D$-forms as a sensitivity of dynamical variables to the 
initial conditions or to the perturbation can be also re-addressed to 
the wave functions. More detailed consideration on the $D$-forms for 
different dynamical realizations in the nonlinear oscillator is considered 
separately elsewhere \cite{iom}.

Originally, a question on sensitivity of wave 
functions to a variance was asked in \cite{perez} to characterize 
quantum chaos by the fidelity of the wave function
\be\label{s3_1}
M(t)=|\lgl\psi_0|\exp\{i\int(\clH+\delta\clH)dt/\hbar\}
\exp\{-i\int\clH dt/\hbar\}|\psi_0\rgl|^2 \, .
\ee
It characterizes an evolution of the initial wave function $\psi_0$ 
governed by the two slightly different Hamiltonians $\clH$ and the 
variational Hamiltonian $ \clH+\delta \clH $.
$M(t)$ was also referred to as the ``Loschmidt echo'' \cite{jpprl01}, 
where dynamics of the initial wave function due to $\clH$ after time $t$ 
is reversed back to the initial state with the variational Hamiltonian 
$\clH+\delta\clH$.
The dynamical decay
of the overlap is characterized by the Lyapunov exponent on some 
semiclasical time scale $\tau_{scl}$ that also stands to be determined.
We will consider analytically the overlap
function $M(t)$ for chaotic dynamics of the nonlinear oscillator 
in the presence of a periodic perturbation of the form of $\delta$--kicks 
with a period $T$ and an amplitude $\ep$. The Hamiltonian of the system is
\be\label{s3_2}
\clH=\hbar\omega a^{\dag}a+\hbar^2\mu(a^{\dag}a)^2-\hbar\ep(a^{\dag}+a)
\sum_{n=-\infty}^{\infty}\delta(t-nT)\equiv\clH_0+V \, .
\ee
 For the initial wave function we choose the coherent states. At the 
moment 
$t=0$ it determined as an eigenfunction of the annihilation operator:
$ a(t=0)|\alpha(t=0)\rgl\equiv a|\alpha\rgl=\alpha|\alpha\rgl $.
It enables one to study their dynamical evolution analytically in the 
semiclassical limit.
Evolution of a generating function was in detail considered in \cite{sok}.
We borrow some details of that analysis with corresponding adaptation to the 
present consideration. The evolution operator is
\be\label{s3_3}
U(t)=\widehat{\exp}\left\{-i\int_0^td\tau\left[\omega a^{\dag}a+
\hbar\mu(a^{\dag}a)^2-\ep g(\tau)(a^{\dag}+a)\right]\right\} \, ,
\ee
where $\widehat{\exp}$ means $T$-ordered exponential. Under this symbol
all exponents commute. Therefore, one can use the following 
Stratonovich--Hubbard transform \cite{habbard}
for the exponential
\be\label{s3_4}
\widehat{\exp}\left[-i\hbar\mu T\int_0^td\tau(a^{\dag}a)^2/T\right]=
\int\prod_{\tau}\frac{d\lambda(\tau)}{\sqrt{4\pi i\kappa}}
e^{i\int_0^td\tau\lambda^2(\tau)/4\kappa}\cdot
e^{-i\int_0^td\tau\lambda(\tau)a^{\dag}a} \, ,
\ee
where we use that $ \kappa=\hbar\mu t $ and $ t/T\rightarrow t $ is a 
number of kicks represented in the continuous form.
We take into account that the harmonic oscillator, acting on the coherent 
state, changes its phase only, and the perturbation $V$ acts as
a shift operator. Namely, {\em e.g.} for the period one writes
\[e^{i\ep(a^{\dag}+a)}e^{-i\phi_{\lambda}(1) a^{\dag}a}
|\alpha\rgl=\exp[i(\ep/2)(\alpha e^{-i\phi_{\lambda}(1)}+
\alpha^*e^{i\phi_{\lambda}(1)}]
|\alpha e^{-i\phi_{\lambda}(1) a^{\dag}a}+i\ep\rgl \, , \]
where $ \phi_{\lambda}(1)=\int_0^1d\tau[\omega T+\lambda(\tau)] $ is a 
phase on 
the period 1. Finely,  we obtain the wave function in the form of the 
following functional integral \cite{sok}
\bea\label{s3_5}
\Psi(t)=U(t)|\alpha\rgl=\int\prod_{\tau}\left(d\lambda(\tau)/
\sqrt{4\pi i\kappa}\right)
\exp\left[i\int_0^td\tau\lambda^2(\tau)/4\kappa\right] \nonumber \\
\times\exp\left[i\ep\int_0^td\tau g(\tau)(\alpha_{\lambda}^*(\tau)+
\alpha_{\lambda}(\tau))/2\right]|\alpha_{\lambda}(t)\rgl \, ,
\eea
where
\be\label{s3_6}
\alpha(t)=e^{-i\phi_{\lambda}(t)}a(t)=
e^{-i\phi_{\lambda}(t)}\left[\alpha+i\ep\int_0^td\tau g(\tau)
e^{i\phi_{\lambda}(\tau)}\right] \, ,
\ee
\be\label{s3_7}
\phi_{\lambda}(t)=\int_0^td\tau[\omega T+\lambda(\tau)] \, .
\ee
The wave function (\ref{s3_5}) is the evolution of the initial ket-vector
$|\alpha\rgl$ due to the Hamiltonian $\clH$ of (\ref{s3_2}).
To obtain an echo, we reverse dynamics  at the moment $t$ back to $t=0$ 
with a random ({\em e.g.} 
Gaussian) time-dependent process $\eta(t)$ to add it to the linear frequency 
$\omega$. 
This variation of the linear frequency affects efficiently the chaos 
control parameter $K$, since the last is $K=4\ep\mu T|\alpha(t)|^2$ \cite{bz}.
Therefore, evolution of the initial bra-vector $\lgl\alpha |$ is due to
the variational Hamiltonian \cite{add2} $\clH+\delta\clH$, with the random 
time-dependent frequency:
\[\omega\rightarrow\omega_{\eta}=\omega+\eta(t)/T \, . \]
The wave function of the variational motion is
\be\label{s3_8}
\Psi_{\eta}^*(t)=\left[U(\eta,t)|\alpha\rgl\right]^{\dag}.
\ee

For simplicity, we consider the Loschmidt echo averaged over the Gaussian 
distribution  
\[P[\eta(t)]\equiv P(\eta(t),\bar{\eta},\sigma)=
\frac{1}{\sqrt{4\pi\sigma}}
\exp\left[-\int_0^td\tau\left(\eta(\tau)-\bar{\eta}\right)^2/4\sigma\right] 
\]
with the non-zero first $\bar{\eta}$ and the second $\sigma$ moments. It 
reads
\be\label{s3_9}
M_{\sigma}(t)=\int\prod_{\tau}d\eta(\tau)P[\eta(\tau)]M(\eta,t) \, ,
\ee
where 
\be\label{s3_10}
M(\eta,t)=|\lgl\Psi_{\eta}(t)|\Psi(t)\rgl|^2
\ee
is the Loschmidt echo for a one fixed realization of $\eta(\tau)$
with $0<\tau<t$. Denoting by
\[\beta_{\lambda}=-i\int_0^td\tau g(\tau)\alpha_{\lambda}(\tau)\, , \]
we obtain the scalar product of the wave function in (\ref{s3_10}) in the 
form
 \bea\label{s3_11}
\clM(\eta,t)\equiv\clM=\lgl\Psi_{\eta}(t)|\Psi(t)\rgl=\int\prod_{\tau}
\frac{d\lambda_1(\tau)d\lambda_2(\tau)}{4\pi\kappa}\exp\left[
\frac{i}{4\kappa}\int_0^td\tau
\left(\lambda_1^2(\tau)-\lambda_2^2(\tau)\right)\right]
\nonumber \\
\times\exp\left[i Im(\alpha_{\lambda_2}^*\alpha_{\lambda_1}-
\ep\beta_{\lambda_1}-\ep \beta_{\lambda_2}^*)-\frac{1}{2}
|\alpha_{\lambda_1}-\alpha_{\lambda_2}|\right].
\eea
In the semiclassical limit $\kappa\ll 1$ this integral is strongly 
simplified and its evaluation is analytically tractable.
After linear change of variables
\be\label{s3_12}
\lambda_1=2\mu+\kappa\nu/2,~~\lambda_2=2\mu-\kappa\nu/2-\eta \, ,
\ee
the phases are
\[ \phi_{\lambda_1}=\int d\tau[\omega T+\lambda_1]=
\int d\tau[\omega T+2\mu+\kappa\nu/2], \]
\[\phi_{\lambda_2}=\int d\tau[\omega T+\lambda_2+\delta]=
\int d\tau[\omega T-2\mu+\kappa\nu/2] .\]
The Jacobian of the transformation is $2\kappa$.
Taking into account these expressions for the phases we obtain from
(\ref{s3_6}) in the semiclassical approximation  that
\be\label{s3_13}
\alpha_{\lambda_2}^*(t)\alpha_{\lambda_1}(t)-\ep\beta_{\lambda_1}-
\ep\beta_{\lambda_2}^*\approx-i\kappa\int_0^td\tau\nu(\tau)|a(\tau)|^2 \, ,
\ee
\be\label{s3_14}
e^{i\phi_{2\mu}(t)}[\alpha_{\lambda_1}-\alpha_{\lambda_2}]\approx
-i\kappa\int_0^td\tau\nu(\tau)a(\tau) \, .
\ee
Here $a(\tau)$, defined in (\ref{s3_6}), is taken for $\nu(\tau)\equiv 0$. 
It gives the definition 
of the classical action as $I(t)=\kappa|a(t)|^2$ for $\kappa\rightarrow 0$
and $|a(t)|^2\rightarrow\infty$.
To carry out integration over $\nu(\tau)$ we use the following auxiliary
expression \cite{sok}
\be\label{auxil}
\exp\left[-(\kappa^2/2)|\int_0^td\tau\nu(\tau)a(\tau)|^2\right]=
\frac{2}{\pi\kappa}\int d^2\xi e^{-2|\xi|^2/\kappa}
\exp\left[-i\frac{\sqrt{\kappa}}{2}Re\xi^*\int_0^td\tau\nu(\tau)a(\tau)\right]
\ee
Substituting (\ref{s3_13}), (\ref{s3_14}) and (\ref{auxil}) into 
(\ref{s3_11}), we obtain for the scalar product of the wave functions
\bea\label{s3_p1}
\clM=\frac{2}{\kappa\pi}\int d^2\xi e^{-2|\xi|^2/\kappa}
\int\prod_{\tau}\frac{d\mu(\tau) d\nu(\tau)}{2\pi}
\exp\left[\frac{-i}{4\kappa}\int d\tau\left(4\mu(\tau)-\eta(\tau)\right)
\left(\kappa\nu(\tau)-\eta(\tau)\right)\right] \nonumber \\
\exp\left[i\kappa\int d\tau\nu(\tau)(|a(\tau)+\xi|^2-|\xi|^2)\right] \, .
\eea
The functional integrations over $\nu(\tau)$ is exact and gives the 
$\delta$--function in $\mu$. Hence the integration over $\mu(\tau)$ is 
also exact. After these integrations we obtain from (\ref{s3_p1})
\be\label{s3_p2}
\clM=\frac{2}{\kappa\pi}\int d^2\xi e^{-2|\xi|^2/\kappa}
\exp\left[\frac{i}{\kappa}\int d\tau\eta(\tau)
\bar{I}_{cl}(\alpha,\alpha^*,\tau)\right] \, ,
\ee
where $\bar{I}_{cl}(\alpha,\alpha^*,\tau)\equiv I(\omega T-2|\xi|^2,
\alpha+\xi,\alpha^*+\xi^*,\tau)$, and we neglect a small term of the 
order of $\eta^2$ in the exponential. Let us expand the last exponential
in (\ref{s3_p2}) in the Taylor series in $\xi$ and $\xi^*$. Therefore,
we have 
\bea\label{s3_p3}
&\exp\left[\frac{i}{\kappa}\int d\tau\eta(\tau)
\bar{I}_{cl}(\alpha,\alpha^*,\tau)\right] \equiv
\clF(\omega T-2|\xi|^2,\alpha+\xi,\alpha^*+\xi^*)\nonumber \\
&=\sum_{m,n,l}\frac{1}{n!m!l!}\frac{\prt^m}{\prt\alpha^m}
\frac{\prt^n}{\prt\alpha^{*n}}\frac{\prt^l}{\prt(\omega T)^l}
\clF(\omega T,\alpha,\alpha^*)\cdot \xi^m\xi^{*n}(-2|\xi|^2)^l \, .
\eea
Substituting (\ref{s3_p3}) in (\ref{s3_p2}) and taking into account that
\[ \frac{2}{\pi\kappa}\int d^2\xi e^{-2|\xi|^2/\kappa}\xi^p\xi^{*q}=
\sqrt{(\kappa/2)^{p+q}}\sqrt{p!q!}\delta_{p,q} \, , \]
we obtain an  expression for the scalar product in the form of the expansion 
in the semiclassical parameter $\kappa$
\be\label{s3_15}
\clM=\sum_{n,l}\frac{(n+l)!}{n!n!l!}
\frac{\prt^{2n}}{\prt\alpha^n\prt\alpha^{*n}}
\frac{\prt^l}{\prt(\omega T)^l}\clF(\omega T,\alpha,\alpha^*)
\cdot (-2)^l(\kappa/2)^{n+l} \, . 
\ee
It should be stressed that the
strongest contribution to the sum (\ref{s3_15}) for the same orders of 
$\kappa$ is due to the derivatives 
over the initial conditions, namely due to the $D$-form: 
\be\label{s3_16}
  D(\tau)\equiv D(I,I)=\frac{1}{\kappa}\left(\frac{\prt 
I(\tau,\alpha,\alpha^*)}{\prt\alpha}\right) \cdot
\left(\frac{\prt I(\tau,\alpha,\alpha^*)}{\prt\alpha^*}\right) 
\propto e^{2\Lambda \tau} \, .   
\ee
Therefore, we obtain, approximately, 
\be\label{s3_17}
\clM_{\eta}(t)\approx\exp\left[\frac{i}{\kappa}\int d\tau \eta(\tau)
I_{cl}(\tau)\right]\exp\left[-\int d\tau \eta(\tau)
\int d\tau'\eta(\tau')D(I_{cl}(\tau),I_{cl}(\tau'))\right] \, .
\ee
Finely, using this approximation, we obtain that the Loschmidt echo
is determined by the following Gaussian integral
\be\label{s3_18}
M_{\sigma}(t)\propto \int\prod_{\tau} d\eta(\tau)
P[\eta(\tau)] |\clM(\eta,t)|^2 \, .
\ee
In the case of narrow packet $\sigma\ll 1$ , that might be roughly 
approximated by a $\delta$-function 
$P[\eta(\tau)]\rightarrow\delta(\eta(\tau)-\bar{\eta})$, we obtain
\be\label{s3_19}
M_{\sigma}(t)\propto \exp\left[-\frac{{\bar{\eta}}^2}{\kappa\Lambda^2}
D(t)\right]\approx\exp\left[-\frac{{\bar{\eta}}^2}{\Lambda^2}
e^{2\Lambda t}\right] \, .
\ee
A possibility of such super-exponential drop off is also mentioned in 
\cite{eckhardt}. For an arbitrary Gaussian distribution, we obtain that
the Loschmidt echo decays exponentially like $\exp\left[-\Lambda 
t\right]$. For instance, in the opposite  case, when $ \sigma\gg 1 $,
using the Fourier transform
\[  
e^{-\kappa A^2(t)}=\int\frac{d\xi}{\sqrt{4\pi}}e^{-\xi^2/4}
e^{-i\xi A(t)} \, ,\]
where $ A(t)=\int_0^td\tau e^{\Lambda\tau}\eta(\tau) $,
we can calculate the functional integral over $\eta(\tau)$ and then over
$\xi$. Finely, we have
\be\label{s3_20}
M_{\sigma}(t)=\left[2\pi+\frac{4\pi\sigma}{\Lambda}
\left(e^{2\Lambda t}-1\right)\right]^{-1/2} \propto 
\sqrt{\frac{\Lambda}{\sigma}}e^{-\Lambda t} \, .
\ee
These decays of $M_{\sigma}(t)$ in (\ref{s3_19}) and (\ref{s3_20}) are 
determined by the classical Lyapunov exponent $\Lambda$.
This result has pure classical nature and is independent of the 
semiclassical parameter $\kappa$, and it survives in the classical
case when $\kappa=0$, as well.

In conclusion,  we presented an  analytical evaluation of the Loschmidt echo
$M(t)$ by means of semiclassical expansion. It leads to 
some restriction on time of the validity of semiclassical description 
which is definitely not coincided with
the quantum breaking time $\thh$ in Eq. (\ref{eqs1}). We shown that $M(t)$ 
has sense on this semiclassical time scale and $M(t)$ decays exponentially 
due to the $D$-form according Eqs. (\ref{s3_17}) and (\ref{s3_19}). 

The validity of Eqs. (\ref{s3_19}) and (\ref{s3_20}) for this time scale 
can be obtained from the semiclassical 
expansion (\ref{s3_13}) and (\ref{s3_14}) in exponential (\ref{s3_11})
that is the semiclassical expansion for the linear oscillator exposed to
the quasi--random field $\lambda$. Therefore, the Jacobian of the 
Hamiltonian flow is
\be\label{s3_21}
J=\det|\prt\left[\alpha(t),\alpha^*(t)\right]/\prt\left[\prt\alpha,\alpha^*
\right]|
\approx
1-2\kappa\sin\phi_{\mu}(t)\int\nu(\tau)d\tau+\kappa^2\left[\int\nu(\tau)d\tau
\right]^2 \, .
\ee
It reflects that the Liouville theorem does not valid for the averages
\cite{biz81}
Since $\nu(\tau)$ is the quasi--random field with the complex Gaussian 
distribution defined at the Stratonovich--Hubbard transform in 
(\ref{s3_4}), we able to evaluate the integral in Eq. (\ref{s3_21}) 
substituting  a mean value of $\nu$. The first moment 
equals zero: $\lgl\nu(\tau)\rgl=0$, that is why we use the second moment 
equaled $\lgl\nu^2(\tau)\rgl=16i/\kappa$. Substituting the 
square root from the modulus of $\lgl\nu^2(\tau)\rgl $ in (\ref{s3_21}),
we obtain for the Jacobian
\[J\approx 1-8\kappa^{1/2}t\sin\phi_{\mu}+16\kappa t^2 \, .\]
Therefore, the validity of the semiclassical expansion carried out in the 
exponential (\ref{s3_11}) and, consequently, validity of the exponential 
decay of $M(t)$ due to the Lyapunov exponent defined in Eq. (\ref{s3_20}) 
is determined by the following semiclassica time scale
\[t<\tau_{scl}\sim 1/4\kappa^{1/2} \, .\]
In the semiclassical limit, when $\kappa\ll 1$ the following inequality
is definitely true: $\tau_{scl}\gg\thh$.
This expression also might be a possible explanation  of the observation
in \cite{jiri} of the exponential decay of $M(t)$ in the Lyapunov regime
for a kicked rotor on times much longer than the Ehrenfest time $\thh$.

I thank P. Gaspard, I. Vanicek, and G.M. Zaslavsky for helpful 
discussions. This research was supported  by the  Minerva Center of 
Nonlinear Physics of Complex Systems.


\begin{thebibliography}{99}
\bibitem{add2} We use here the variational Hamiltonian 
with the variance $\delta\clH=\delta(t)a^{\dag}a/T$ instead of the
perturbed  one to not confuse it with the time dependent 
perturbation $V$.
\end{thebibliography}
\end{document}